\input{aipcheck}

\documentclass[
    ,final            % use final for the camera ready runs
  ]
  {aipproc}

\layoutstyle{6x9}

\newcommand{\bce}{\begin{center}} 
\newcommand{\ece}{\end{center}}
\newcommand{\beq}{\begin{equation}}
\newcommand{\eeq}{\end{equation}}
\newcommand{\bea}{\vspace{0.25cm}\begin{eqnarray}}
\newcommand{\eea}{\end{eqnarray}}

\newcommand{\br}{{\bf r}}

\newcommand{\ba}{\begin{array}}
\newcommand{\ea}{\end{array}}

\newcommand{\ket}[1]{| {#1} \rangle}

\def\lsim{\mathrel{\rlap{\lower4pt\hbox{\hskip1pt$\sim$}}
    \raise1pt\hbox{$<$}}}         %less than or approx. symbol
\def\gsim{\mathrel{\rlap{\lower4pt\hbox{\hskip1pt$\sim$}}
    \raise1pt\hbox{$>$}}}         %greater than or approx. symbol

\def\beq{\begin{equation}}
\def\endeq{\end{equation}}
\def\arr{\begin{eqnarray}}
\def\endarr{\end{eqnarray}}

\begin{document}

\title{Current non-conservation effects in $\nu$DIS
diffraction}


\author{R.~Fiore}{
  address={Dipartimento di Fisica,
Universit\`a     della Calabria
and
 Istituto Nazionale
di Fisica Nucleare, Gruppo collegato di Cosenza,
I-87036 Rende, Cosenza, Italy}
}

\author{ V.R.~Zoller}{
  address={ITEP, Moscow 117218, Russia}
}

\begin{abstract}
In the  neutrino DIS diffraction  the charged  current non-conservation 
gives rise to sizable  corrections to the longitudinal 
structure function,
 $F_L$. 
 These corrections is a  higher twist effect
 enhanced at small-$x$ by the  rapidly growing  gluon density.
The phenomenon 
manifests 
itself in abundant  
  production  of  charm and strangeness by   longitudinally 
polarized W bosons 
of moderate virtualities $Q^2\lsim m_c^2$
\end{abstract}

\maketitle

%%%%%%%%%%%%%%%%%%%%%%%%%%%%%%%%%%%%%%%%%%%%
%% MAINMATTER
%%%%%%%%%%%%%%%%%%%%%%%%%%%%%%%%%%%%%%%%%%%%

\section{Introduction}
Weak currents are not conserved. Here we focus
on manifestations of the charmed-strange ($cs$) 
charged current  non-conservation (CCNC)
in small-x neutrino DIS. For light flavors the hypothesis of the 
partial conservation
 of the axial-vector current (PCAC) \cite{PCAC} quantifies 
the  CCNC in terms of observable quantities 
\cite{Adler}.  The $cs$ current  non-conservation is not 
constrained by PCAC and we quantify the $cs$CCNC in terms of 
the light cone wave functions of the color dipole QCD approach.  
The observable   highly  
sensitive to the 
CCNC effects is  the 
so called longitudinal structure function $F_L(x,Q^2)$. Our finding is that
the  higher twist  correction to $F_L$ arising  from  the 
$cs$CCNC   appears to be  
enhanced at small $x$ by the BFKL \cite{BFKL} gluon density factor,
\beq
F_L^{cs} \sim {m_c^2\over Q^2}\left(1\over x\right)^{\Delta}.
\label{eq:HT}
\eeq 
As a result, the component of  $F_L(x,Q^2)$ induced by the charmed-strange 
current grows rapidly to small-$x$ and 
 dominates $F_L$ at   $Q^2\sim m_c^2$ \cite{FZCS1,FZCS2}.

\section{CCNC in terms of LCWF }
In the color dipole (CD) approach to small-$x$ 
$\nu$DIS \cite{Kolya92} the responsibility for the quark current
 non-conservation
 takes the light-cone wave function (LCWF) of the quark-antiquark
Fock state of the longitudinal ($L$)  electro-weak boson.
If the Cabibbo-suppressed transitions are neglected, 
the Fock state  expansion 
  reads
\beq
\ket{W^+_L}=\Psi^{cs}\ket{c\bar s}
+\Psi^{ud}\ket{u\bar d}+...,
\label{eq:FOCK}
\eeq  
where only   $u\bar d$- and $c\bar s$-states (vector and axial-vector) 
are retained.
 
In  the current conserving eDIS the Fock state expansion  of 
the longitudinal photon
   contains only $S$-wave $q\bar q$ states and $\Psi$  vanishes as  $Q^2\to 0$,
\beq
\Psi(z,{\bf r})\sim 
2\delta_{\lambda,-\bar\lambda}Q z(1-z)\log(1/\varepsilon r).
\label{eq:V1}
\eeq

In $\nu$DIS the CCNC  adds to Eq.(\ref{eq:V1})
 the $S$-wave  mass term \cite{FZ1,FZ2}
\beq
\sim \delta_{\lambda,-\bar\lambda}Q^{-1}
\left[(m\pm \mu)[(1-z)m\pm z \mu]\right]\log(1/\varepsilon r)
\label{eq:V2}
\eeq
and generates the $P$-wave component of $\Psi(z,{\bf r})$, 
\beq
\sim i\delta_{\lambda,\bar\lambda}e^{-i2\lambda\phi}Q^{-1}(m\pm \mu)r^{-1}
\label{eq:P1}
\eeq
 (upper sign - for the  axial current, lower - for the vector one).
 Clearly seen are the built-in
divergences of
the vector and axial-vector currents
$\partial_\mu V^{\mu}\sim m-\mu$ and  $\partial_\mu A^{\mu}\sim m+\mu$. 
This LCWF describes the quark antiquark state with  quark  of mass  
$m$ and  helicity   $\lambda=\pm 1/2$
carrying fraction $z$ of the $W^+$ light-cone momentum and 
antiquark having  mass $\mu$, helicity   $\bar\lambda=\pm 1/2$ and 
momentum fraction $1-z$.
The distribution of  dipole sizes, $r$, is controlled by the 
attenuation parameter
$$\varepsilon^2=Q^2z(1-z)+(1-z)m^2+z\mu^2$$ 
that introduces the 
infrared cut-off, 
$r^2\sim\varepsilon^{-2}$.  

\section{High $Q^2$: $z$-symmetric $c\bar s$-states}
In the color dipole representation \cite{NZ91,M}
the  longitudinal structure 
function $F_L(x,Q^2)$ 
in  the vacuum exchange dominated  region
of $x\lsim 0.01$ 
can be represented in a factorized form
\beq
F_{L}(x,Q^{2})
={Q^2\over 4\pi^2\alpha_{W}}\int dz d^{2}{\bf{r}}
|\Psi(z,{\bf{r}})|^{2} 
\sigma(x,r)\,,
\label{eq:FACTOR}
\eeq
where $\alpha_W=g^2/4\pi$ and 
${G_F/\sqrt{2}}={g^2/m^2_{W}}$. The light 
cone density of 
color  dipole states  $|\Psi|^2$ 
is the incoherent sum of the vector $(V)$  and the  
axial-vector $(A)$ terms,
$$|\Psi|^2= |V|^2+ |A|^2$$.

\begin{figure}
  \includegraphics[height=.3\textheight]{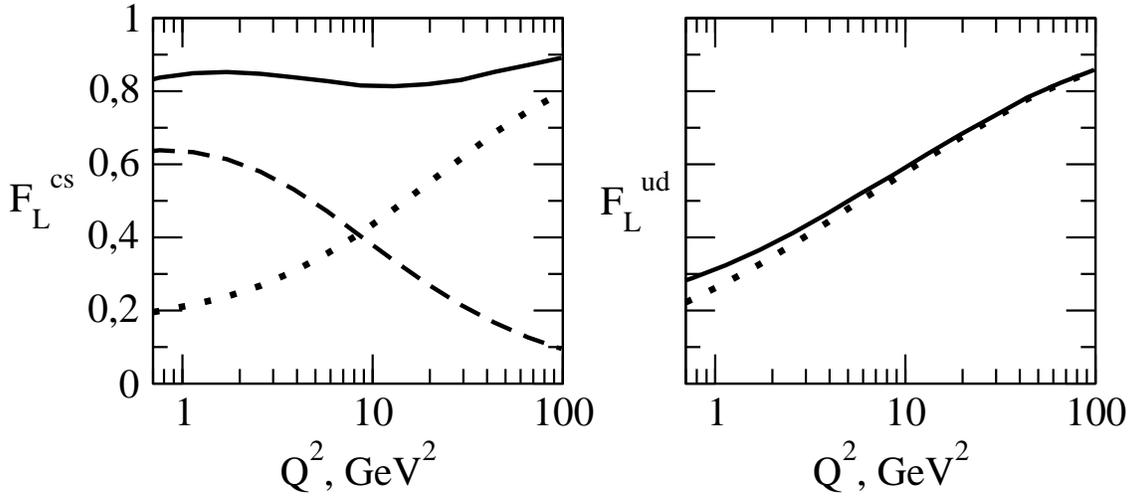}
  \caption{Two components of 
$F_L=F^{cs}_L + F^{ud}_L$ at $x_{Bj}=10^{-4}$ are shown 
by solid lines. The $S$-wave and $P$-wave contributions to $F^{cs}_L$  and 
$F^{ud}_L$ are represented by dotted and dashed lines, correspondingly.}
\label{fig:fig1}
\end{figure}

The Eqs. (\ref{eq:V1},\ref{eq:FACTOR}) make it evident
that  for large enough virtualities of the probe, $Q^2\gg m_c^2$,
   the $ S$-wave  components of 
\beq
F^{(\nu)}_L=F^{ud}_L+F^{cs}_L
\label{eq:FLUDCS}
\eeq
 corresponding  to the ``non-partonic'' 
configurations with  $z\sim 1/2$ do dominate \cite{BGNPZ94} and 
two terms in the   
 expansion (\ref{eq:FLUDCS})
that mimics the expansion (\ref{eq:FOCK}) do   converge 
(see Fig.~\ref{fig:fig1}).
To the Double Leading Log approximation (DGLAP \cite{D,GLAP})
\beq
F^{ud}_L\approx F^{cs}_L\approx {2\over 3\pi}\alpha_S(Q^2)G(x,Q^2).
\label{eq:FL2FL}
\eeq
The $rhs$ of (\ref{eq:FL2FL})  is quite similar to  $F^{(e)}_L$ of 
eDIS \cite{D,Cooper} 
(see \cite{NZFL} for discussion of corrections to 
Double Leading Log-relationships 
between the gluon density $G$ and $F^{(e)}_L$). 
\section{Moderate $Q^2$: asymmetric $c\bar s$-states and $P$-wave dominance}
The $S$-wave term dominates $F_L$  at high 
$Q^2\gg m_c^2$.
At $Q^2\lsim m_c^2$ the $P$-wave component takes over 
(see Fig.{\ref{fig:fig1}}). To  evaluate it 
we turn to  Eq. (\ref{eq:FACTOR}).
For   $m_c^2\gg m_s^2$, 
$$
|V_L|^2\sim|A_L|^2\propto
 ({m_c^2/Q^2})
\varepsilon^2 K^2_1(\varepsilon r)$$
 and
corresponding  $z$-distribution, $dF^{cs}_L/dz$,  
 develops  
the parton model peaks at $z\to 0$ and $z\to 1$ \cite{FZCS1}. 
Integrating  over  $z$  near the endpoint $z=1$ in (\ref{eq:FACTOR})  
yields \cite{FZCS2}
\bea
\int dz |\Psi^{cs}(z,\br)|^2\approx {\alpha_WN_c\over \pi^2}{m_c^2\over
 m_c^2+Q^2}{1\over Q^2r^4}
\label{eq:ZINT1}
\eea
for $r^2$ from 
$
(m_c^{2}+Q^2)^{-1}\lsim r^2 \ll m_s^{-2}$
This is the $r$-distribution for $c\bar s$-dipoles with $c$-quark
 carrying a fraction 
$z\sim 1$ of 
the  $ W^+$'s  light-cone momentum.
 
The lowest  order pQCD cross section  \cite{NZ91}
$$\sigma(r)\approx \pi C_F\alpha_S^2r^2\log\left(1/r^2\right)$$
 saturates for large dipoles and can be approximated by   
$$
\sigma(r)\approx \pi C_F\alpha_S^2r^2\log\left(1+r_s^2/r^2\right).$$
The saturation radius  is found to be
$r_S^2={A/ \mu_G^2},$
where  $A\simeq 10$ \cite{NZFL} and $\mu_G=1/R_c$ is the inverse 
correlation radius of perturbative gluons.  
 From the lattice QCD  studies $R_c\simeq 0.2-0.3$ fm \cite{MEGGIO}. 
Then, for the charmed-strange $P$-wave component of  $F_L$ with fast $c$-quark
($z\to 1$) one gets
\beq
F_L^{cs}\approx {N_cC_F\over 8}{m_c^2\over {m_c^2+Q^2}}
\left(\alpha_S\over \pi\right)^2
\log^2\left[(Q^2+m_c^2)r_S^2\right].
\label{eq:FLZ1}
\eeq

Additional contribution to $F^{cs}_L$ 
comes from the $P$-wave $c\bar s$-dipoles 
with ``slow'' $c$-quark,  $z\to 0$.
 For low $Q^2\ll m_c^2$ this contribution is rather small,
\beq
F_L^{cs}\approx {N_cC_F\over 4}
{Q^2+m_s^2\over m_c^2}\left(\alpha_S^2\over\pi\right)^2\log(r_S^2m_c^2).
\label{eq:FLZ0}
\eeq
If, however,  $Q^2$ is large enough, $Q^2\gsim  m_c^2$,
corresponding distribution of dipole sizes
\bea
\int dz |\Psi^{cs}(z,\br)|^2\approx {\alpha_WN_c\over \pi^2}{m_c^2\over
 m_s^2+Q^2}{1\over Q^2r^4}
\label{eq:ZINT2}
\eea
 valid for 
$ 
(m_c^{2}+Q^2)^{-1}\lsim r^2 \ll m_c^{-2}$  and  $z\to 0$
leads to 
\beq
F_L^{cs}\approx {N_cC_F\over 8}{m_c^2\over {Q^2}}\left(\alpha_S\over \pi\right)^2
\log^2\left({Q^2+m_c^2\over m_c^2}\right),
\label{eq:FL2G}
\eeq
Therefore, at high $Q^2\gg m_c^2$ both  kinematical 
domains  $z\to 0$ and $z\to  1$ (Eqs.(\ref{eq:FL2G}) and (\ref{eq:FLZ1}), 
respectively)
contribute equally to $F_L^{cs}$  and one can anticipate similar 
 $x$-dependence of both 
contributions. 

In the CD  approach the BFKL-$\log(1/x)$ evolution  
of  $\sigma(x,r)$
is described by the CD BFKL equation of Ref.\cite{NZZBFKL}. 
For qualitative estimates it suffices to use  the DGLAP approximation. 
The DGLAP resummation results in
the $P$-wave component of $F_L$  that rises rapidly to small $x$,
\bea
F_L^{cs}\approx  {N_cC_F\over 2}{m_c^2\over {Q^2}}{L(Q^2)\eta(x)^{-1}}
I_2\left(2\sqrt{\xi(x,Q^2)}\right).
\label{eq:DGLAP}
\eea
In Eq.(\ref{eq:DGLAP}),  which is the DGLAP-counterpart of Eq.(\ref{eq:HT}), 
$I_2(z)\simeq \exp(z)/\sqrt{2\pi z}$ 
is the Bessel function, 
$$\xi(x,Q^2)= \eta(x) L(Q^2)$$ is the DGLAP expansion parameter with
$$
L(k^2)={4\over \beta_0}
\log[{\alpha_S(\mu_G^2)/\alpha_S(k^2)}],$$
$$\alpha_S(k^2)={4\pi\over \beta_0}\log(k^2/\Lambda^2)$$ and
$
\eta(x)=C_A\log(x_0/x).$
%%%%%%%%%%%%%%%%%%%%%%%%%%%%%%%%%%%%%%%%%%%%
%% Sample figure:
%%
%% The option [height=...] scales the picture to the given height,
%% without it it would be printed at its nominal size
%%%%%%%%%%%%%%%%%%%%%%%%%%%%%%%%%%%%%%%%%%%%

As for our numerical estimates (Fig.~\ref{fig:fig1}),  
we calculate nuclear and nucleon structure functions to the 
leading order in 
$\alpha_S\log{(1/x)}$ within the 
color dipole BFKL approach  \cite{NSZZ}. The  full scale BFKL evolution of 
$F_L(x,Q^2)$ is shown in Fig. 2 of Ref.\cite{FZAdler}. 

\section{Summary} 
 
Summarizing, it is shown  that at small $x$ and moderate   
virtualities of the probe, $Q^2\sim m_c^2$, the higher twist corrections
brought about by  the non-conservation  of the  charmed-strange current 
dramatically change the longitudinal structure function, $F_L$.
   The effect   survives 
the limit $Q^2\to 0$ and  
seems to be interesting from a point of view of 
 feasible  tests of Adler's theorem \cite{Adler} and the PCAC hypothesis.

\section{Acknowledgments}
The work was supported in part by the Ministero Italiano
dell'Istruzione, dell'Universit\`a e della Ricerca and  by
 the RFBR grant 06-02-16905  and 07-02-00021.



%%%\bibliographystyle{aipproc}   % if natbib is available

%%%\bibliography{sample}

\IfFileExists{\jobname.bbl}{}
 {\typeout{}
  \typeout{******************************************}
  \typeout{** Please run "bibtex \jobname" to optain}
  \typeout{** the bibliography and then re-run LaTeX}
  \typeout{** twice to fix the references!}
  \typeout{******************************************}
  \typeout{}
 }

%\end{document}

%%%%%%%%%%%%%%%%%%%%%%%%%%%%%%%%%%%%%%%%%%%
%% The following lines show an example how to produce a bibliography
%% without the help of the BibTeX program. This could be used instead
%% of the above.
%%%%%%%%%%%%%%%%%%%%%%%%%%%%%%%%%%%%%%%%%%%

\end{document}